\documentclass[11pt]{article}
\usepackage{hyperref}
\usepackage{graphicx}
\pdfoutput=1
\begin{document}
\title{Superfast Thinning of a  \\ Nanoscale Thin Liquid Film}
\author{Michael Winkler$^{1)}$, Guggi Kofod$^{1)}$, Rumen Krastev$^{3)}$, and Markus Abel$^{1),2)}$\\
\\\vspace{6pt}
1) Institute for Physics and Astronomy, University of Potsdam, 14476 Potsdam,\\
2) Universit\'e Henri Poincar\'e-LEMTA, BP 160 - 54504 Vandoeuvre, \\
3) NMI - Natural and Medical Science Institute, 72770 Reutlingen,
}

\maketitle

\begin{abstract}
This fluid dynamics video demonstrates an experiment on superfast thinning of a freestanding
thin aqueous film. The production of such films is of fundamental interest for interfacial sciences
and the applications in nanoscience. The stable phase of the film is of the order $5-50\,nm$; nevertheless
thermal convection can be established which changes qualitatively the thinning behavior from
linear to exponentially fast. The film is thermally driven on one spot by a very cold needle,
establishing two convection rolls at a Rayleigh number of $10^7$. This in turn enforces thermal
and mechanical fluctuations which change the thinning behavior in a peculiar way, as shown in the
video.
\end{abstract}

\begin{figure*}
  \includegraphics[width=\textwidth, draft=false]{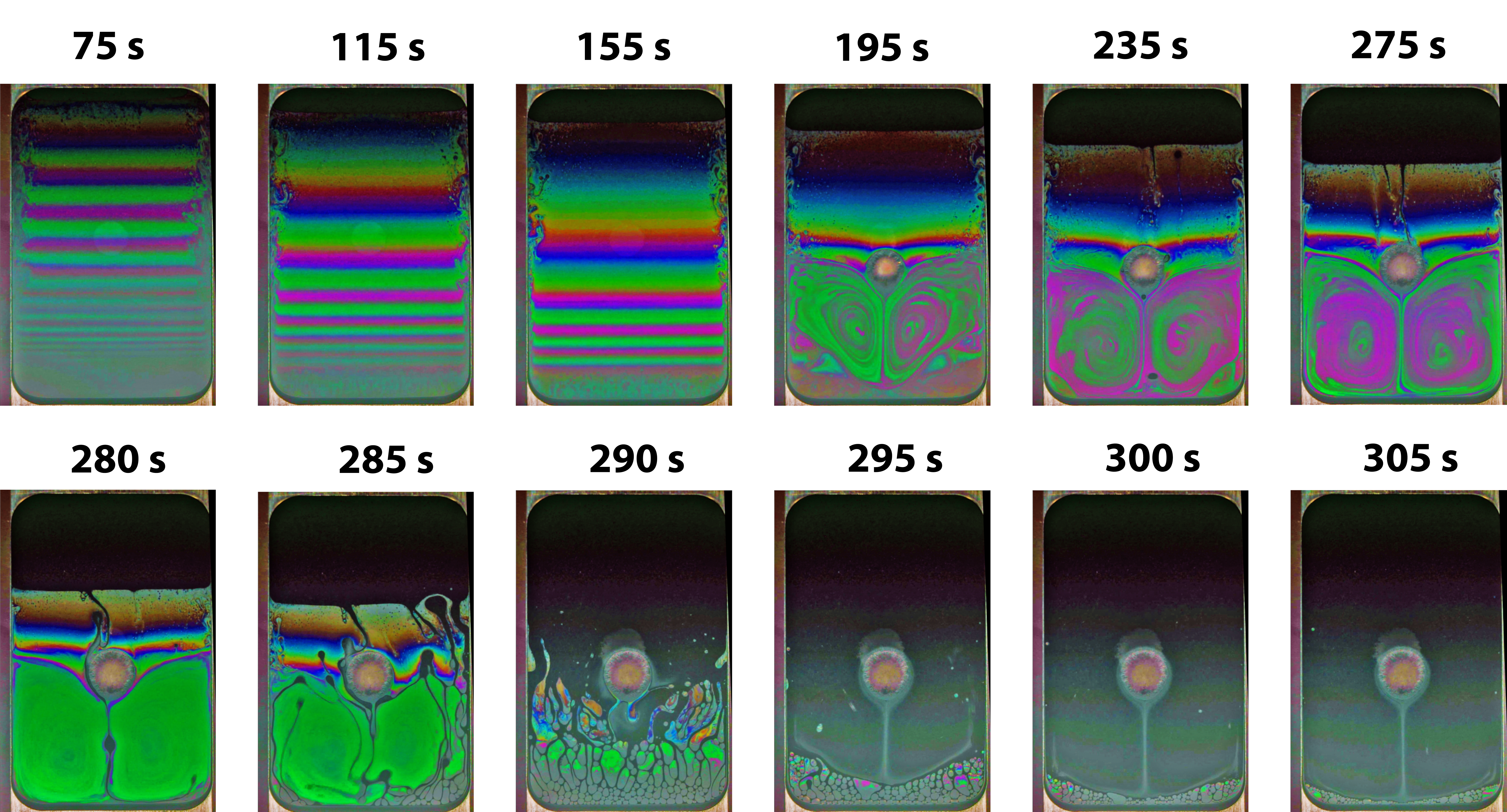}
\caption{Thinning by turbulent convection.
We show the film evolution for some dynamically interesting instants of time.
While the usual thinning evolves slowly, convection
mixes and transports material around very quickly
until the regime of bubble convection is reached.
The involved processes contain convection, turbulent mixing, surface forces,
disjoining pressure and gravitation. The black area on the top of each
picture contains optically invisible material of thickness $<50\,nm$. The colors
correspond to different thicknesses, corresponding to a negative interference
of wavelength $n\cdot \lambda/4 =d$, i.e. a certain color can occur for several
thicknesses.
The time step in the first row is $40 \,s$ and $5 \,s$ in the second row.}
\label{fig:thinningsequence}
\end{figure*}

Thin liquid films may show a very thin stable phase of nanometer scale.
Since the film is no longer visible by optical wavelengths at this thickness,
it is called a Black Film \cite{derjaguin1989theory}. The evolution of an
initially thick, freestanding film
towards this equilibrium thickness is a slow process which can be observed as a flat
boundary of Black Film on top of a periodic color pattern moving downwards.
The colors correspond to the repeated negative interference of light waves
when the condition $n\cdot \lambda/4$ is met. This process is driven by gravitation and
surface forces, but the time scale is set by the Poiseuille flow between the
film interfaces \cite{couder1989hydrodynamics}, cf. Fig.~\ref{fig:thinningsequence}, 75s and 115s.

When the thin film is driven, the motion may change due to the altered transport properties.
We explore this possibility by thermal driving with a cold copper rod at $100 \,K$, corresponding
to a Rayleigh number of $10^7$.
This establishes two stable convection rolls, Fig.~\ref{fig:thinningsequence}, 195s--275s
and gives rise to large mechanical and
thermal fluctuations. These fluctuations in turn generate spontaneously spots of
stable and light black film inside the unstable thick and heavy phase.
The spots are convected for small size until they eventually escape to the top
due to buoyancy. While being convected, the spots grow and leave behind tails of
Black Film, thereby increasing the Black Film area in an exponential manner,
cf. Fig.~\ref{fig:thinningsequence}, $280\,s--305\,s$.

\bibliographystyle{plain}
\bibliography{gallery}

\end{document}